\documentclass[12pt]{article}
\usepackage{graphicx}
\usepackage{amsmath}
\usepackage{amssymb}
\usepackage{caption2}
\begin{document}
\bibliographystyle{prsty}
\begin{center}
{\large {\bf \sc{  Final-state interactions in the decay $B^0 \to \eta_c K^*$  }}} \\[2mm]
Zhi-Gang Wang \footnote{ E-mail,wangzgyiti@yahoo.com.cn.  }    \\
 Department of Physics, North China Electric Power University, Baoding 071003, P. R. China
\end{center}

\begin{abstract}
In this article, we study the  final-state rescattering effects in
the decay $B^0 \to \eta_cK^*$, the numerical results indicate the
corrections are comparable with the  contribution from the naive
factorizable amplitude, and the total amplitudes can accommodate
the experimental data.
\end{abstract}

PACS numbers:  13.20.He; 14.40.Lb

{\bf{Key Words:}}  Final-state interactions, $B$-decay
\section{Introduction}
 The nonleptonic decays of  the $B$ meson have attracted much
attention in studying the
 nonperturbative dynamics of QCD, final-state interactions and CP violation.
The exclusive  $B$  to  charmonia decays  are of great importance
since the decays $B \to J/\psi K, \eta_c K, \chi_{cJ} K $ are
regarded as the golden channels for  studying  CP violation. The
quantitative understanding of those decays depends  on our knowledge
about the nonperturbative hadronic matrix elements of the operators
entering the effective weak Hamiltonian.  The $B$ factories (BaBar,
Belle, etc) have measured the color-suppressed decays  to a
charmonium and a $K$ (or $K^*$) meson with relatively large
branching fractions \cite{PDG}, for examples, $\rm{Br}(B^0\to J/\psi
K^0)=(8.72\pm0.33)\times 10^{-4}$, $\rm{Br}(B^+\to J/\psi
K^+)=(10.07\pm0.35)\times 10^{-4}$, $\rm{Br}(B^0\to \eta_c
K^0)=(9.9\pm1.9)\times 10^{-4}$, $\rm{Br}(B^+\to \eta_c
K^+)=(9.1\pm1.3)\times 10^{-4}$, $\rm{Br}(B^0\to \eta_c
K^{*0})=(1.6\pm0.7)\times 10^{-3}$, etc,  which take place  through
the process $b \to sc \bar{c}$ (or more precise $\bar{b} \to \bar{s}
c\bar{c}$, they relate with each other by  charge conjunction, in
this article, we calculate the amplitudes for the process $b \to sc
\bar{c}$, then take  charge conjunction to obtain the branching
fraction.) at the quark-level.

Recently, the BaBar Collaboration  measured  the processes
$B^0\to\eta_c K^{*0}$, $B^0\to\eta_c(2S) K^{*0}$, $B^0\to h_c
K^{*0}$, $B^+\to h_c K^+$ and $\eta_c(2S)\to K\bar{K}\pi$ from the
$B$-decays to $(K\bar{K}\pi)K^+$, $(K\bar{K}\pi)K^{*0}$,
$\eta_c\gamma K^+$ and $\eta_c\gamma K^{*0}$. The branching
fractions are $\rm{Br}(B^0\to\eta_c K^{*0})=(5.7\pm0.6 \pm0.9
)\times 10^{-4}$,
 $\rm{Br}(B^0\to\eta_c(2S) K^{*0})<3.9\times 10^{-4}$, $\rm{Br}(B^+\to h_c K^+)
 \rm{Br}(h_c\to\eta_c\gamma)<4.8\times 10^{-5}$ and $\rm{Br}(B^0\to h_c K^{*0}) \rm{Br}(h_c\to\eta_c\gamma)<2.2\times
 10^{-4}$ at the $90\%$ C.L. \cite{BABAR0804}.

The decays $B\rightarrow J/\psi K,\eta_cK, \chi_{cJ}K$ have been
calculated with the QCD-improved factorization approach
\cite{BBNS1,BBNS2},  there are infrared divergence in vertex
corrections and logarithmical divergence in spectator corrections
beyond the leading twist approximation for the $S$-wave charmonia
and  in the leading twist approximation for the $P$-wave charmonia,
moreover, the predicted  branching fractions are too small to
accommodate the experimental data
\cite{ChayF,ChengF,ChaoF1,ChaoF2,ChaoF3,PhamF}. The decay
$B\rightarrow J/\psi K^*$ has also been studied  with the
QCD-improved factorization approach, the factorization breaks down
even at the twist-2 level for transverse hard spectator interactions
 \cite{ChengFV}.

In Refs.\cite{MelicLC1,WangLC,MelicLC2}, the soft nonfactorizable
contributions in the decays  $B\to J/\psi K,
\chi_{c0}K,\chi_{c1}K,\eta_cK$ are studied with the light-cone QCD
sum rules, the predicted  small branching fractions  cannot
accommodate the (relatively large)  experimental data.

 In Ref.\cite{pQCD}, the authors study the decays $B\to J/\psi K$, $\chi_{c0}K$, $\chi_{c1}K$, $\eta_cK$,
  $J/\psi K^*$,
 $\chi_{c0}K^*$, $\chi_{c1}K^*$, $\eta_cK^*$ with the perturbative QCD approach based on $k_T$ factorization
 theorem, and  the branching fractions $\rm{Br}(B \to \eta_cK)$ and $\rm{Br}(B^0\to\eta_c
K^{*0})$ ($2.64^{+2.71}_{-2.58}\times 10^{-4}$) are too small to
take into account the  experimental data \cite{BABAR0804}.

Final-state interactions play an important role in the hadronic
$B$-decays,  the  color-suppressed neutral modes such as $B^0 \to
D^0 \pi^0, \pi^0\pi^0, \rho^0\pi^0, K^0\pi^0$ are  enhanced
substantially by the long-distance rescattering effects \cite{CHHY}.
In Refs.\cite{ColangeloFSI1,ColangeloFSI2}, the authors study  the
rescattering effects of the  intermediate charmed mesons for the
decays $B^- \to \chi_{c0}K^- , h_cK^- $
  and observe the final-state interactions can
lead to larger branching fractions to account the experimental data.
The factorizable amplitude in the decay  $B^0\to \eta_cK^*$ is too
small to accommodate the experimental data \cite{BABAR0804}, so it
is intersecting to study the effects of the final-state
interactions.

The article is arranged as: in Section 2, we study  the final-state
rescattering effects in the decay  $B^0 \to \eta_cK^* $; in Section
3, the numerical result and discussion; and Section 4 is reserved
for conclusion.

\section{Final-state rescattering effects in the decay  $B^0 \to \eta_cK^* $}

The effective weak Hamiltonian for the  decay modes $b\rightarrow s
c \bar{c}$  can be written as (for detailed discussion of the
effective weak Hamiltonian, one can consult Ref.\cite{Buras})
\begin{equation}
H_w = \frac{G_F}{\sqrt{2}}\left\{V_{c b} V_{c s}^* \left[ C_1(\mu)
{\cal O}_1(\mu) +  C_2(\mu) {\cal O}_2(\mu) \right] -V_{tb}
V_{ts}^*\sum_{i=3}^{10} C_i(\mu){\cal O}_i(\mu) \right\}  \, ,
\end{equation}
where $V_{ij}$'s are the CKM matrix elements, $C_i$'s are the Wilson
coefficients calculated at the renormalization scale $\mu \sim
O(m_b)$ and the relevant operators ${\cal O}_i$  are given by
 \begin{eqnarray}
 {\cal O}_1&=&(\overline{s}_{\alpha} b_{\alpha})_{V-A}
(\overline{c}_{\beta} c_{\beta})_{V-A}\, , \nonumber\\
 {\cal O}_2&=&(\overline{s}_{\alpha} b_{\beta})_{V-A}
(\overline{c}_{\beta} c_{\alpha})_{V-A}\,,
 \nonumber\\
 {\cal O}_{3(5)}&=&(\overline{s}_{\alpha} b_{\alpha})_{V-A} \sum_q
(\overline{q}_{\beta} q_{\beta})_{V-A(V+A)}\,,
 \nonumber\\
  {\cal
O}_{4(6)}&=&(\overline{s}_{\alpha} b_{\beta})_{V-A}  \sum_q
(\overline{q}_{\beta} q_{\alpha})_{V-A(V+A)}\,,
 \nonumber\\
{\cal O}_{7(9)}&=&{3\over 2}(\overline{s}_{\alpha} b_{\alpha})_{V-A}
\sum_q e_q (\overline{q}_{\beta} q_{\beta})_{V+A(V-A)}\,,
 \nonumber\\
  {\cal O}_{8(10)}&=&{3\over 2}(\overline{s}_{\alpha} b_{\beta})_{V-A}
\sum_q e_q (\overline{q}_{\beta} q_{\alpha})_{V+A(V-A)} \, ,
 \end{eqnarray}
 $\alpha$ and $\beta$ are color indexes. We can reorganize  the
color-mismatched quark fields into color singlet states by Fierz
transformation (for example, $ {\cal {O}}_2 = \frac{1}{3} {\cal
{O}}_1 + 2 {\cal \widetilde{O}}_1
 $, ${\cal \widetilde{O}}_1=(\overline{s} \frac{\lambda^a}{2} b )_{V-A}
(\overline{c}\frac{\lambda^a}{2} c)_{V-A}$,  $\lambda^a$'s are the
 Gell-Mann matrices),   and obtain the factorizable
amplitude,
\begin{eqnarray}
\langle \eta_c(p_3)K^*(p_4)| H_w|B(P) \rangle &=&
\frac{G_F}{\sqrt{2}} \left\{V_{c b} V_{c s}^*
(C_1+\frac{C_2}{3})-V_{tb} V_{ts}^*(C_3-C_5+\frac{C_4-C_6}{3})
\right\} \nonumber \\
&& \langle \eta_c(p_3)|\overline{c} \gamma_{\mu}(1-\gamma_5) c|0
\rangle\langle K^*(p_4)|\overline{s} \gamma^{\mu}
(1-\gamma_5)b|B(P) \rangle \nonumber \\
&=&\frac{G_F}{\sqrt{2}} \left\{V_{c b} V_{c s}^*
(C_1+\frac{C_2}{3})-V_{tb} V_{ts}^*(C_3-C_5+\frac{C_4-C_6}{3})
\right\} \nonumber \\
&&2P\cdot\epsilon^*_4 f_{\eta_c}M_{K^*}A_0(p_3^2) \, ,
\end{eqnarray}
 where we have used the standard definitions for the weak
form-factors (we write down all form-factors to be used in this
article) \cite{Form1,Form2},
\begin{eqnarray}
\langle D(p)| \overline{s} \gamma_{\mu}(1-\gamma_5) b | B(P) \rangle
&=& (P + p)_{\mu}  F_{1}(q^2) - \frac{ M_B^2 - M_D^2 }{q^2} q_{\mu}
[F_{1}(q^2) - F_0(q^2)]\, , \nonumber\\
\langle V(p)| \overline{s} \gamma_{\mu}(1-\gamma_5) b | B(P)
\rangle&=&i{\epsilon_\mu}^{\nu\alpha\beta}\epsilon^*_\nu P_\alpha
p_\beta
\frac{2V(q^2)}{M_B+M_{V}} -\frac{2M_{V}q\cdot\epsilon^*}{q^2}q_\mu A_0(q^2)\nonumber\\
&&-\left[\epsilon_\mu^*-\frac{q\cdot\epsilon^*}{q^2}q_\mu\right](M_B+M_{V})A_1(q^2)+\nonumber\\
&&\left[(P+p)_\mu-\frac{M_B^2-M_{V}^2}{q^2}q_\mu\right]q\cdot\epsilon^*
\frac{A_2(q^2)}{M_B+M_{V}} \, ,
\end{eqnarray}
the $\epsilon_\mu$ is the polarization vector of the vector meson
and $q_\mu=P_\mu-p_\mu$. In this article, we use the value of the $B
\to K^*$ form-factor $A_0(q^2)$ from the light-cone QCD sum rules
\cite{BallV},
\begin{eqnarray}
A_0(q^2)&=&\frac{1.364}{1-q^2/M_B^2}-\frac{0.990}{1-q^2/36.78} \, .
\end{eqnarray}
 The  factorizable amplitude (see
Eq.(3)) at the tree level is too small to accommodate the
experimental data.

The decays $B^0 \to DD_s$, $DD_s^*$, $D^*D_s$, $D^*D_s^*$  are color
enhanced due to the large Wilson coefficient $C_2$,
\begin{eqnarray}
\langle D_s(q)D^*(p)| H_w|B(P) \rangle
&=&\sqrt{2}G_F
V_{c b} V_{c s}^* (C_2+\frac{C_1}{3})P\cdot\epsilon^* f_{D_s}M_{D^*}A_0(q^2)\, ,\nonumber \\
\langle D_s^*(q)D(p)| H_w|B(P) \rangle &=&\sqrt{2}G_FV_{c b} V_{c
s}^* (C_2+\frac{C_1}{3}) P\cdot\epsilon^* f_{D_s^*}M_{D^*_s}F_1(q^2)
\, ,
\end{eqnarray}
we write down only the amplitudes appear in the final expressions.
In the heavy quark limit, the weak form-factors $A_0(q^2)$ and
 $F_1(q^2)$ can be related to the universal Isgur-Wise
form-factor $\xi(\omega)$ \cite{HeavyQuark},
\begin{eqnarray}
F_1(q^2)&=&\frac{M_B+M_D}{2\sqrt{M_BM_D}}\xi\left(\frac{M_B^2+M_D^2-q^2}{2M_BM_D}\right)
\, ,\nonumber \\
A_0(q^2)&=&\frac{M_B+M_{D^*}}{2\sqrt{M_BM_{D^*}}}\xi\left(\frac{M_B^2+M_{D^*}^2-q^2}{2M_BM_{D^*}}\right)
\, ,
\end{eqnarray}
where $\xi(\omega)=\left(\frac{2}{1+\omega}\right)^2$, which is
compatible with the experimental data  for the semileptonic decays
$B\to D^*(D)l\nu_l$ \cite{IS}.

The decay $B^0 \to \eta_c K^* $ can also take place through the
decay cascades $B^0 \to DD_s$, $DD_s^*$, $D^*D_s$, $D^*D_s^*$ $\to
\eta_c K^*$, the rescattering amplitudes of  $DD_s$, $DD_s^*$,
$D^*D_s$, $D^*D_s^*$ $\to \eta_c K^*$ may play an important role.

The final-state interactions can be described  by the following
effective lagrangians,
\begin{eqnarray}
\mathcal{L}_{\eta_c D^*D}&=&-i g_{\eta_c D^*D} \eta_c
\left[\partial^\mu D{D}_\mu^{*\dagger}- {D}_\mu^{*} \partial^\mu{D}^{\dagger}\right]\, ,\\
\mathcal{L}_{\eta_c D^*D^*}&=& g_{\eta_c D^*
D^*}\varepsilon^{\mu\nu\alpha\beta} \eta_c \partial_\mu {D}_\nu^{*\dagger} \partial_\alpha {D}_\beta^{*} \, ,\\
\mathcal{L}_{DDV}&=&-ig_{DDV}D_{i}^{\dagger}{\stackrel{\leftrightarrow}{\partial}}
_{\mu}D^{j}(\mathbb{V}^{\mu})^{i}_{j} \, ,\\
\mathcal{L}_{D^*DV}&=&-2f_{D^{*}DV}\varepsilon^{\mu\nu\alpha\beta}(\partial_{\mu}\mathbb{V}_{\nu})^{i}_{j}
\left[ D_{i}^{\dagger}{\stackrel{\leftrightarrow}{\partial}}_{\alpha}D_\beta^{* j}-D^{*\dagger}_{\beta i}{\stackrel{\leftrightarrow}{\partial}}_{\alpha}D^{j}\right]\, ,\\
\mathcal{L}_{D^*D^*V}&=&ig_
{D^{*}D^{*}V}D^{*\nu\dagger}_{i}{\stackrel{\leftrightarrow}{\partial}}_{\mu}
D^{*j}_{\nu}(\mathbb{V}^{\mu})^{i}_{j}+4if_{D^{*}D^{*}V}D^{*\dagger}_{i\mu}(\partial^{\mu}\mathbb{V}^{\nu}
-\partial^{\nu}\mathbb{V}^{\mu})^{i}_{j}D^{*j}_{\nu}\, ,
\end{eqnarray}
where the indexes $i,j$  stand for  the flavors of the light quarks,
$D^{(*)}$=$(\bar{D}^{(*)0}$, $D^{(*)-}$, $D_{s}^{(*)-})^T$,
$\mathbb{V}$ is the $3\times 3$ matrix for the nonet vector mesons,
\begin{eqnarray}
\mathbb{V}&=&\left(\begin{array}{ccc}
\frac{\rho^{0}}{\sqrt{2}}+\frac{\omega}{\sqrt{2}}&\rho^{+}&K^{*+}\\
\rho^{-}&-\frac{\rho^{0}}{\sqrt{2}}+\frac{\omega}{\sqrt{2}}&
K^{*0}\\
K^{*-} &\bar{K}^{*0}&\phi
\end{array}\right).
\end{eqnarray}
 The lagrangians
$\mathcal{L}_{DDV}$, $\mathcal{L}_{D^*DV}$ and
$\mathcal{L}_{D^*D^*V}$ are taken from Ref.\cite{CHHY}, and the
$\mathcal{L}_{\eta_c D^*D}$ and $\mathcal{L}_{\eta_c D^*D^*}$ are
constructed from the   heavy quark theory  in this article. In the
heavy quark limit, the strong coupling constants $f_{D^*DV}$,
$f_{D^*D^*V}$, $g_{DDV}$ and $g_{D^*D^*V}$ can be related to the
basic parameters  $\lambda$ and $\beta$ in the heavy quark effective
Lagrangian (one can consult Ref.\cite{HQEFT97} for the heavy  quark
effective lagrangian and relevant parameters,  we neglect them for
simplicity),
\begin{eqnarray}
 f_{D^*DV}&=&\frac{f_{D^*D^*V}}{M_{D^*}}=\frac{\lambda g_V}{\sqrt{2}}\, ,\nonumber\\
g_{DDV}&=&g_{D^*D^*V}=\frac{\beta g_V}{\sqrt{2}}\, ,
\end{eqnarray}
where  $g_V=5.8$  from the vector meson dominance theory
\cite{VMDgV}. The strong coupling constants $g_{\eta_c D^*D}$ and
$g_{\eta_c D^*D^*}$ are estimated with the  universal Isgur-Wise
form-factor at zero recoil $\xi(1)$  and the  assumption of
dominance of the intermediate $\eta_c$ meson for the pseudoscalar
heavy quark current $\overline{c}i\gamma_5 c$,
\begin{eqnarray}
g_{\eta_c D^*D}&=&\frac{2m_c}{f_{\eta_c}}\approx
\frac{M_{\eta_c}}{f_{\eta_c}} \, , \nonumber \\
g_{\eta_c D^*D^*}&=& \frac{g_{\eta_c D^*D}}{M_{D^*}} \, .
\end{eqnarray}

The rescattering effects can be taken into account  by twelve
Feynman diagrams,  see Fig.1. We  calculate the absorptive parts (or
imaginary parts) of the rescattering amplitudes $\textbf{Abs}(i)$ by
the Cutkosky rule,
$\textbf{Abs}(i)=\widetilde{\textbf{Abs}}(i)\frac{G_F}{\sqrt{2}}
V_{c b} V_{c s}^* (C_2+\frac{C_1}{3})$,
\begin{eqnarray}
\widetilde{\textbf{Abs}}(a)&=&\frac{|\vec{p}_1|}{8\pi^2M_B}\int
d\Omega\left[f_{D_s}M_{D^*} g_{\eta_c D^*D}
g_{DDV}A_0(p_2^2)\right]\frac{\mathcal{F}^2(M_{D},t)}{t-M_D^2}\nonumber\\
&& p_2\cdot\epsilon_4^* p_2\cdot\epsilon_1^* q\cdot\epsilon_1
 \, ,\nonumber\\
\widetilde{\textbf{Abs}}(b)&=&\frac{|\vec{p}_1|}{8\pi^2M_B}\int
d\Omega\left[f_{D_s^*}M_{D_s^*} g_{\eta_c D^*D}
g_{DDV}F_1(p_1^2)\right]\frac{\mathcal{F}^2(M_{D_s},t)}{t-M_{D_s}^2}\nonumber\\
&& p_2\cdot\epsilon_4^* p_2\cdot\epsilon_1^* q\cdot\epsilon_1
 \, ,\nonumber\\
\widetilde{\textbf{Abs}}(c)&=&-\frac{|\vec{p}_1|}{4\pi^2M_B}\int
d\Omega\left[f_{D_s}M_{D^*} g_{\eta_c D^*D^*}
f_{D^*DV}A_0(p_2^2)\right]\frac{\mathcal{F}^2(M_{D^*},t)}{t-M_{D^*}^2}\nonumber\\
&&\epsilon^{\mu\nu\alpha\beta}\epsilon^{\mu'\nu'\alpha'\beta}p_{4\mu}\epsilon_{4\nu}
 p_{2\alpha}p_{1\mu'}p_{2\nu'}q_{\alpha'}
 \, ,\nonumber\\
 \widetilde{\textbf{Abs}}(d)&=&-\frac{|\vec{p}_1|}{4\pi^2M_B}\int
d\Omega\left[f_{D_s^*}M_{D^*_s} g_{\eta_c D^*D^*}
f_{D^*DV}F_1(p_1^2)\right]\frac{\mathcal{F}^2(M_{D^*_s},t)}{t-M_{D^*_s}^2}\nonumber\\
&&\epsilon^{\mu\nu\alpha\beta}\epsilon^{\mu'\beta\alpha'\beta'}p_{4\mu}\epsilon_{4\nu}
 p_{2\alpha}q_{\mu'}p_{1\alpha'}p_{2\beta'}
 \, ,\nonumber\\
  \widetilde{\textbf{Abs}}(e)&=&\frac{|\vec{p}_1|}{8\pi^2M_B}\int
d\Omega\left[f_{D_s^*}M_{D^*_s} g_{\eta_c D^*D}F_1(p_2^2)\right]\frac{\mathcal{F}^2(M_{D^*},t)}{t-M_{D^*}^2}\nonumber\\
&&\left[-g_{D^*D^*V} p_2 \cdot \epsilon_4^* p_1 \cdot
\epsilon^*(q)\epsilon(q)\cdot\epsilon_2 p_1 \cdot \epsilon^*_2
\right.\nonumber\\
&&-2f_{D^*D^*V}p_4 \cdot \epsilon_2 p_1 \cdot \epsilon^*_2
p_1\cdot\epsilon^*(q)\epsilon(q)\cdot\epsilon^*_4
 \nonumber\\
 &&\left.+2f_{D^*D^*V}p_4 \cdot \epsilon(q) p_1 \cdot \epsilon^*(q)
p_1\cdot\epsilon^*_2\epsilon_2\cdot\epsilon^*_4 \right]
 \, ,\nonumber\\
 \widetilde{ \textbf{Abs}}(f)&=&\frac{|\vec{p}_1|}{8\pi^2M_B}\int
d\Omega\left[f_{D_s}M_{D^*} g_{\eta_c D^*D}A_0(p_1^2)\right]\frac{\mathcal{F}^2(M_{D^*_s},t)}{t-M_{D^*_s}^2}\nonumber\\
&&\left[-g_{D^*D^*V} p_2 \cdot \epsilon_4^* p_1 \cdot
\epsilon^*(q)\epsilon(q)\cdot\epsilon_2 p_1 \cdot \epsilon^*_2
\right.\nonumber\\
&&-2f_{D^*D^*V}p_4 \cdot \epsilon_2 p_1 \cdot \epsilon^*_2
p_1\cdot\epsilon^*(q)\epsilon(q)\cdot\epsilon^*_4
 \nonumber\\
 &&\left.+2f_{D^*D^*V}p_4 \cdot \epsilon(q) p_1 \cdot \epsilon^*(q)
p_1\cdot\epsilon^*_2\epsilon_2\cdot\epsilon^*_4 \right] \, ,
\nonumber \\
\widetilde{ \textbf{Abs}}(g)&=& -\widetilde{ \textbf{Abs}}(h )
\,\,\text{ (in SU(3) limit})\, ,
\nonumber \\
\widetilde{ \textbf{Abs}}(i)&=& -\widetilde{ \textbf{Abs}}(j
)\,\,\text{ (in SU(3) limit}) \, ,
\nonumber \\
\widetilde{ \textbf{Abs}}(k)&=& -\widetilde{ \textbf{Abs}}(l )
\,\,\text{ (in SU(3) limit})\, ,
\end{eqnarray}
where $\vec{p}_1$ is the 3-momentum of the on-shell intermediate
mesons $D ,D^*, D_s, D_s^*$   in the rest frame of the $B$ meson,
for example, in the process   $B^0 \to D^*(p_1)D_s(p_2)\to
\eta_c(p_3)K^*(p_4)$, $t=q^2$, $q=p_1-p_3=p_4-p_2$, $\epsilon^\mu_i$
is the polarization vector of the  vector meson $i$. The off-shell
effects of the $t$-channel exchanged  mesons $D$, $D^*$, $D_s$ and
$D_s^*$ are  taken into account  by introducing a monopole
form-factor \cite{CHHY},
\begin{eqnarray}
\mathcal{F}(M_{i},t)=\frac{\Lambda_i^{2}-M_{i}^2 }{\Lambda_i^{2}-t},
\end{eqnarray}
and the cutoff $\Lambda_i$ are parameterized as
\begin{eqnarray}
\Lambda_{i}=M_{i}+\alpha \Lambda_{\rm{QCD}} \, ,
\end{eqnarray}
where $\alpha$ is a free parameter and
$\Lambda_{\rm{QCD}}=0.225\rm{GeV}$. In fact, the
$g_s\mathcal{F}(M_{i},t)$ are the momentum dependent strong coupling
constants, we can vary the parameter $\alpha$ to change the
effective strong couplings,  here we use the notation $g_s$ to
denote all the strong coupling constants.

The dispersive parts (or real parts) of the rescattering amplitudes
can be obtained via the dispersion relation,
\begin{eqnarray}
 \textbf{Dis}(i) (M_B^2) = {1 \over \pi} \textbf{P}
\int_{s_{th}}^{\infty} { \textbf{Abs}(i) (s^\prime) \over s^\prime -
M_B^2} d s^\prime \, ,
\end{eqnarray}
where the thresholds $s_{th}$ are given by $s_{th}=(
M_D+M_{D^*_s})^2$ , $( M_{D^*}+M_{D_s})^2$, $( M_D+M_{D_s})^2$, $(
M_{D^*}+M_{D^*_s})^2$ for any specific diagram. There are large
uncertainties due to the cut-off procedure, even  one assume that
the integrals are dominated by the region close to the pole $M_B^2$
\cite{ColangeloFSI1,ColangeloFSI2}. In this article, we assume the
dominating contributions of the rescattering amplitudes come from
the absorptive parts,  which originate from the on-shell
intermediate states in the decay cascades, the dispersive parts of
the amplitudes are of minor importance and can be taken into account
by introducing a phenomenological parameter $\rho$, $
\textbf{Dis}(i)=\rho \textbf{Abs}(i)$, $\rho\leq 30\%$.

\begin{figure}
\centering
  \includegraphics[totalheight=16cm,width=12cm]{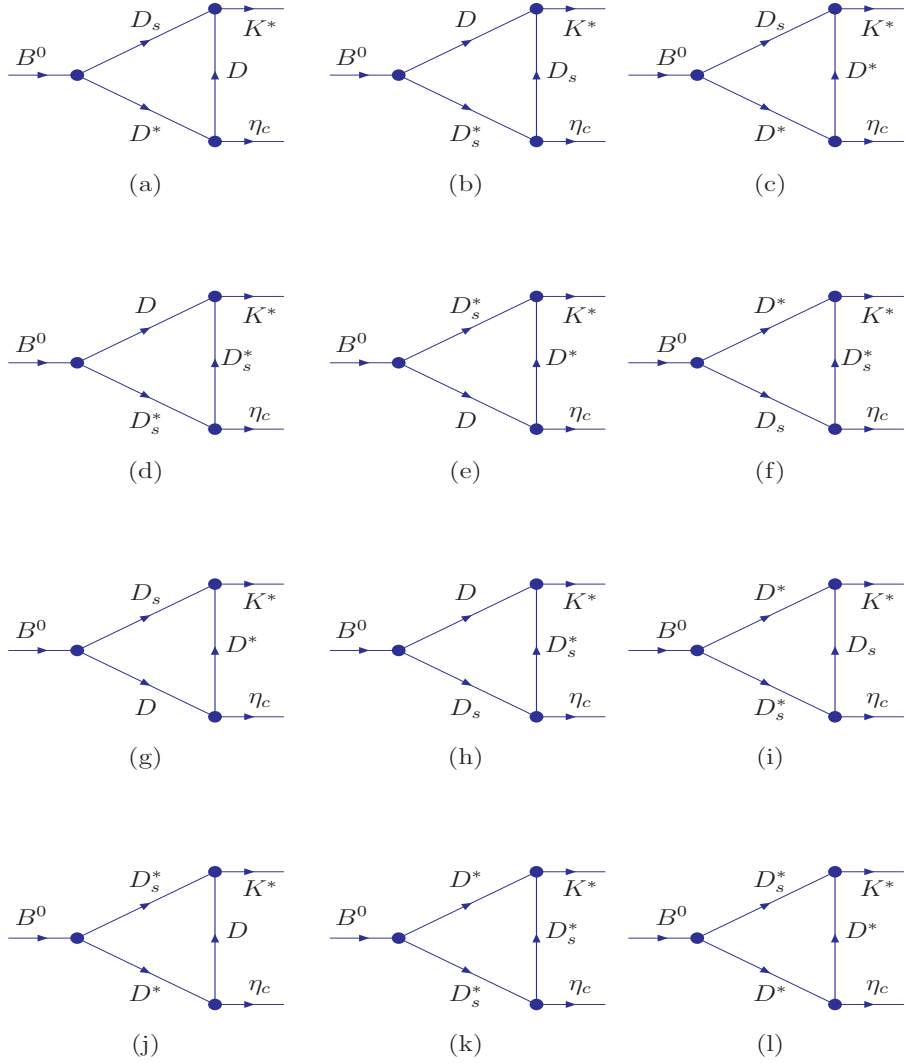}
      \caption{The Feynman diagrams for the final-state interactions. }
\end{figure}

\section{Numerical result and discussions}
The CKM matrix elements  are taken as $V_{cs}=0.97296\pm0.00024$,
$V_{cb}=(41.6\pm0.6)\times 10^{-3}$,
$V_{tb}=0.999100^{+0.000034}_{-0.000004}$ and
$V_{ts}=-(40.6\pm2.7)\times 10^{-3}$ \cite{PDG,CKM}. We take the
next-to-leading order Wilson coefficients calculated in the naive
dimensional regularization scheme
 for $\mu = \overline{m_b}(m_b) = 4.40 {\rm \, GeV}$ and
$\Lambda_{\overline{\rm MS}}^{(5)} = 225\, {\rm MeV}$, $ C_1(\mu) =
-0.185$,  $ C_2(\mu) = 1.082$, $C_3(\mu) = 0.014$,  $C_4(\mu) =
-0.035 $,  $C_5(\mu) = 0.009 $  and  $C_6(\mu) = -0.041 $
\cite{Buras}, here we have neglected the Wilson coefficients
$C_7,C_8,C_9,C_{10}$ in numerical calculation due to their small
values. The masses of the mesons are taken as $M_B = 5.279\, \rm{
GeV}$, $M_{K^*}=0.892\,\rm{GeV}$,
 $M_{D}=1.87\,\rm{GeV}$,
$M_{D_s}=1.97\,\rm{GeV}$, $M_{D^*}=2.010\,\rm{GeV}$ and
$M_{D_s^*}=2.112\,\rm{GeV}$ \cite{PDG}, and
$M_{\eta_c}=2.986\,\rm{GeV}$ \cite{BABAR0804}.

The values of the decay constants $f_D$, $f_{D_s}$, $f_{D^*}$ and
$f_{D^*_s}$ vary in a large range from different approaches, for
examples, the potential model, QCD sum rules and lattice QCD, etc
\cite{decayC1,decayC2,decayC3}. For the $f_D$, we take  the
experimental data from the CLEO Collaboration,
$f_D=222.6\pm16.7^{+2.8}_{-3.4}\,\,\rm{MeV}$
\cite{decayCP1,decayCP2}. The value
$f_{D_s}=(0.274\pm0.013)\,\rm{GeV}$ from the CLEO Collaboration
 shows the $SU(3)$ breaking effect is rather large \cite{decayCPs},
$\frac{f_{D_s}}{f_D}=1.23$, while most of theoretical calculations
indicate $\frac{f_{D_s}}{f_D}\approx 1.1$,  we take the value
$f_{D^*}=f_D=(0.22\pm0.02)\,\rm{GeV}$ and
$f_{D^*_s}=f_{D_s}=(0.24\pm0.02)\,\rm{GeV}$.

The decay constant $f_{\eta_c}$ can be estimated with the QCD sum
rules \cite{SVZ} or phenomenological potential models,  the values
from those approaches are compatible with each other, we can take
the value $f_{\eta_c} = (0.35\pm0.02) \,\rm{ GeV}$
\cite{WangSD,constant,PRT1978}.

The basic parameters $\lambda$ and $\beta$ in the heavy quark
effective Lagrangian are estimated with the vector meson dominance
theory \cite{VMD1,VMD2},  $\lambda=0.56\,\rm{GeV}^{-1}$ and
$\beta=0.9$. The corresponding values  of the strong coupling
constants are
\begin{eqnarray}
 f_{D^*DV} &=&2.30 \,\rm{GeV}^{-1} \, ,\nonumber \\
 f_{D^*D^*V} &=&4.61  \, ,\nonumber \\
 g_{DDV} &=&3.69 \, , \nonumber \\
 g_{D^*D^*V} &=&3.69\, ,
\end{eqnarray}
while the values from the light-cone QCD sum rules are much smaller
\cite{WangCCT1,WangCCT2}. In this article, the strong coupling
constants $g_{\eta_c D^*D}$ and $g_{\eta_c D^*D^*}$ are estimated
with the universal Isgur-Wise form-factor at zero recoil $\xi(1)$
and the assumption of dominance of the intermediate $\eta_c$ meson
for the pseudoscalar heavy quark current $\overline{c}i\gamma_5 c$.
We take the results from the vector meson dominance theory   for
consistence. However, we may overestimate the final-state
rescattering effects due to the larger strong coupling constants,
and have to compensate them with suitable
 $\alpha$.

The parameters $\alpha$ and $\rho$ are taken to be $\rho=0.3$,
$\alpha=1.4-1.8$. The contributions from the rescattering effects
are somewhat sensitive to the parameter $\alpha$ (or the constant
$\Lambda_i$ in the form-factors),  the $\Lambda_i$ is of the order
of the mass of radial excitations of the charmed mesons
\cite{ColangeloFSI1,ColangeloFSI2}.

 Finally  we obtain the numerical results for the branching fractions,
\begin{eqnarray}
\rm{Br}(B^0 \to \eta_c K^*)&=& (3.25-4.09) \times 10^{-4} \,\,\,\text{(Tree amplitude)}\, ,\nonumber \\
\rm{Br}(B^0 \to \eta_c K^*)&=& (3.90-5.31)\times 10^{-4}
\,\,\,\text{(Tree+Abs
amplitude)} \, ,\nonumber \\
\rm{Br}(B^0 \to \eta_c K^*)&=& (4.83-6.94)\times 10^{-4}
\,\,\,\text{(Tree+Abs+Dis
 amplitude)}\, ,\nonumber \\
 \rm{Br}(B^0\to\eta_cK^{*})&=&(5.7\pm0.6 \pm0.9 )\times 10^{-4} \,\,\,\text{(Experimental
 data)} \, .
\end{eqnarray}

 \section{Conclusion}
In this article, we study the  final-state rescattering effects in
the decay $B^0 \to \eta_cK^*$, the numerical results indicate the
corrections are comparable with the  contribution from the naive
factorizable amplitude, and the total amplitudes can accommodate the
experimental data.

\section*{Acknowledgments}
This  work is supported by National Natural Science Foundation,
Grant Number  10775051, and Program for New Century Excellent
Talents in University, Grant Number NCET-07-0282.

\end{document}